\documentclass[12pt]{article}

\def\pd{\partial_{\mu}}
\def\pa{\partial}
\def\pu{\partial^{\mu}}
   
\def\p{\phi}

\def\l{\lambda}
\def\m{\mu}
\def\n{\nu}   
\def\e{\epsilon}
\def\s{\sigma}

\def\g{\gamma}

\def\a{\alpha}
  
\def\r{\rho}
\def\vp{\varphi}

\def\F{{\cal F}}

\def\A{{\cal A}}
\def\be{\begin{equation}}
\def\ee{\end{equation}}
\def\ba{\begin{eqnarray}}
\def\ea{\end{eqnarray}}
\def\z{{ }^{*}}
   
\def\h{\hat}  
  
\def\f0{f_0}  
\begin{document}
\thispagestyle{empty}
\vspace{3cm}
\rightline{SINP-2001-37/677}
\vspace{4cm}
\centerline{\large\bf  String-Loop Corrected Magnetic Black Holes }
\vspace{2cm}
\centerline{\bf Mikhail Z. Iofa \footnote{iofa@theory.sinp.msu.ru}}
\centerline{Skobeltsyn Institute of Nuclear Physics}  
\centerline{Moscow State University}
\centerline{Moscow 119899, Russia}
\centerline{\today}
\vspace{2cm}

\begin{abstract}
We discuss  the form of the string-loop-corrected effective action
obtained by compactification of the heterotic
string theory on the manifold $K3\times T^2$ or on its orbifold limit and
the loop-corrected magnetic black hole solutions of the equations of
motion. Effective 4D theory has  N=2 local supersymmetry.
Using the string-loop-corrected prepotential of the N=2 supersymmetric
theory,
which receives corrections only from the string world sheets of torus
topology, we calculate the loop corrections
to the tree-level gauge couplings  and solve the loop-corrected
equations of motion.  At the string-tree level, the
effective gauge couplings decrease at small distances from the origin,
and in this region
string-loop corrections to the  gauge couplings become important.
A possibility of smearing  the singularity of the
tree-level supersymmetric solution with partially broken supersymmetry
 by quantum corrections is discussed.
\end{abstract}
\pagebreak
\section{Introduction}
At present, string theory is considered the best candidate for
 a fundamental theory that would provide a consistent quantum theory of
gravity unified with the other interactions \cite{gsw}.
In particular, string theory provides a powerful approach to the physics of
black holes (review and further refs. in \cite{yo,peet}). In this setting, 
we meet a fundamental problem  of
understanding how the intrinsically stringy effects modify  the Einstein
gravity. 

In this paper we discuss two of these effects:  presence of
scalar fields such as the dilaton and  moduli, and higher-genus
contributions modifying  the tree-level effective action. 
We focus on higher-genus corrections, because 
string theory, being a theory formulated on a world sheet, always contains
string-loop corrections from higher topologies of the world sheet (for the
two-derivative terms these
vanish  for higher supersymmetries $N\geq 4$), while ${\a}'$ corrections 
can vanish  in  certain constructions based on conformal field theories and 
for a large class of backgrounds \cite{bh1,bh2}.  

We consider the perturbative 
 one-string-loop (torus topology) corrections for a special class of 
backgrounds: $4D$ black holes provided by  the "chiral null
models" \cite{bh1,bh2} embedded in heterotic string theory 
compactified on the manifold $K3\times T^2$.
For this compactification pattern, the effective $4D$ field theory is
described by $N=2$ supergravity interacting with matter.

The universal sector contains supergravity and vector multiplets, the vector
components resulting from dimensional reduction of $6D$ metric and the
antisymmetric tensor on the two-torus $T^2$.  By using
the (perturbative) one-loop-corrected prepotential in the $N=2$
supergravity, we calculate the effective gauge couplings in the universal
sector of the theory. Due to the $N=2$ supersymmetry, there are no 
loop corrections
to the prepotential beyond the one loop \cite{afgnt,wikalu,defekoz}.

In string perturbation theory, higher order
contributions enter with the factor
$e^{-{1\over2}\chi {\p'}}$, where $\chi$ is the Euler characteristic of 
the string world sheet.
The exponent $e^{\p_\infty}\equiv \e $, where $\p_\infty 
\equiv\p|_{|x|\rightarrow\infty}$  can be considered as a string-loop
expansion parameter~\footnote{For closed surfaces without boundaries, 
$\chi =2-2g$ is
the number of handles. The factor $e^{\p_\infty} $ multiplying the
tree-level  effective action
is absorbed in the Newton coupling constant.}. The dilaton $\p'$ is split 
as $\p + \p_\infty $.

In the case of magnetic black hole with the charges $P_1$ and $P_2$ 
 the string-tree-level dilaton 
 $\p =\frac{1}{2}\ln\frac{(r+P_{1} )(r+P_{2})}{r^2} $ increases, and the 
tree-level
gauge couplings of the vector fields in the universal sector which are
proportional to $e^{-\p}$ decrease at small distances, so that the effective 
gauge couplings are sensitive to  string-loop corrections. 

The loop-corrected configuration of magnetic black hole can be obtained
 by solving either the equations of motion derived from the loop-corrected
effective action, or spinor Killing equations (conditions for supersymmetry
variations of spinors to vanish) ( review and refs. in \cite{moh}). 

In this paper we follow the first way, i.e. solve the loop-corrected 
equations of motion. Solution of spinor Killing
equations is presented elsewhere \cite{mi}. 
Equations of motion are solved  analytically in the first order in the
loop-counting parameter $\e$. Considerable technical simplification 
is achieved for a special choice
of  magnetic charges, in which case the the tree-level metric of the
two-torus is independent of coordinates.  However, qualitatively, the results 
remain unchanged for unequal charges.

Solving the system of the equations of motion for the moduli
and the Einstein-Maxwell equations
in the first order in $\e$, we obtain the loop-corrected metric and dilaton. 
A family of solutions for the metric  of the loop-corrected
magnetic black hole is
$$ g_{ii}=-g^{00} = \left(1+\frac{P}{r}\right) +\e
\left(A_1 \frac{P}{r}-A_2 \frac{P}{r+P}\right),
$$
where $A_i$ are arbitrary constants, $P=8\sqrt{P^1 P^2}$. 
Supersymmetry fixes the value of $A_i$:
$A_1=-\frac{Ph}{2TU},\,\, A_2=0$, 
where $h(T,U)$ is the real  (unambiguous) part of the 
one-string-loop correction to the prepotential, $T,U$ are the
standard string tree-level moduli. For a particular compactification
of the heterotic string for which the loop correction
to the prepotential was calculated it appears to be  strictly negative
\cite{hm}.
The above metric can be considered as the first term in the expansion in the
loop-counting parameter of the metric
$$ g_{ii}=-g^{00} = 1+\frac{P}{r +\e\frac{r+|\e PH|}{r+P}}. 
$$ 
When extrapolated to the region of small $r$, this  metric
has no singularity, which is smeared by quantum corrections.

In Sect.2 we review the structure of the 4D magnetic black hole
in heterotic string theory provided by  chiral null model.

In Sect.3 we discuss the $N=2$ locally supersymmetric effective theory.

Using the symplectic structure of the theory, we calculate the gauge
couplings in a basis admitting the prepotential and by symplectic
transformation obtain them in the basis associated with the heterotic string
compactification (in which case the prepotential does not exist)
and construct  the loop-corrected action. 

In Sect.4 we solve the loop-corrected equations of motion. Starting from
the tree-level extremal magnetic black-hole solution, in the first order in
string coupling constant, we
obtain an analytic solution of the loop-corrected equations of motion.

Solution of the loop-corrected equations of
motion and calculation of the loop-corrected gauge couplings is
presented in the Appendixes.  
\section{Charged 4D black hole solutions in 4D effective field theory.}
The effective field theory which describes dynamics of the light
fields in four dimensions depends on  the pattern of compactification of
the initial superstring theory. Heterotic string theory, when compactified 
on the manifold $K3$ or on a suitable orbifold 
yields locally supersymmetric $6D$ theory with $N=1$ supersymmetry,
 while compactification of the heterotic string on the 4-torus $T^4$
results in  $6D$, $N=2 $  locally supersymmetric theory.

The bosonic part of the universal sector 
 for both compactifications has the same form
\begin{equation}
\label{F1}
I_6=\frac{1}{2{\kappa_6}^2}\int d^6 x\sqrt{-G}e^{-{\Phi}'}\left 
[R + (\partial {\Phi}'
)^2 - \frac{H^2}{12} \right ]+\ldots.
\end{equation}

A large class of solutions to the equations of motion following from the
action (\ref{F1}) is provided by the chiral null models \cite{bh1,bh2,howe}.
From the string-theoretical point of view, a chiral null model
is a conformal 2D theory interpreted as string world-sheet 
Lagrangian with nontrivial backgrounds:
\begin{equation}
   L =  F(x)  \pa u \left [\bar{\pa} v +
   K(u,x)\bar{\pa}  u  +   2{\cal A}_i(u,x)  \bar{\pa}  x^i
\right]
 +   (G_{ij} + B_{ij})(x) \pa x^i \bar{\pa} x^j    +    {\cal R}
\Phi (x)\  .
 \label{E1}
\end{equation}
As solutions of the equations of motion of the full effective theory,
 the backgrounds ensure vanishing of the $\beta$-functions 
 and conformal invariance of the corresponding 2D theory. 

An important special case of  chiral null models (\ref{E1}) is provided by 
 the chiral null models with  curved transverse part of the form
\cite{bh1,bh2,howe}  
\begin{eqnarray}
 L = F(x)  \partial u \bigg(\bar{\partial} v +  K(x) \bar{\partial} u\bigg)  +
 f(x)k(x)  \big[ \partial x^4 + a_s (x) \partial x^s\big] \big[
\bar{\pa} x^4  + a_s (x) \bar{\pa} x^s\big]\nonumber \\
    + f(x)  k^{-1} (x) \partial x^s \bar{\partial} x^s
 +   b_s (x) (\partial x^4 \bar{\partial} x^s - \bar{\partial} x^4 \partial
x^s)
  +   {\textstyle {1\over 8}} \alpha'
\sqrt{ g^{(2)}}  R^{(2)}   \ln F(x)f(x), 
 \label{E2}
\end{eqnarray}
Here $x^s=(x^1,x^2,x^3)$ and $v=2t$ are  non-compact space-time coordinates,
$u=y^2$ and  $x^4=y^1$ are compact toroidal coordinates. 
 Written in $4D$ form, the Lagrangian of the chiral null model is
$$
L =   (G'_{\mu\nu} + B'_{\mu\nu}) (x) \partial x^\mu \bar{\partial} x^\nu
+
 G_{mn}(x) [ \partial y^m   + A^{(1)m}_{\mu}(x)  \partial x^\mu]
  [ \bar{\partial} y^n   + A^{(1)n}_{\nu}(x)  \bar{\partial} x^\nu]
$$
\begin{equation}
+ \ A^{(2)}_{n\mu}(x)  (\partial y^n \bar{\partial} x^\mu - \bar{\partial}
y^n\partial x^\mu)
+  {\textstyle {1\over 8}} \alpha'
\sqrt{ g^{(2)}}  R^{(2)} \Phi'(x)  .
\label{E3}
\end{equation}

The background fields are independent of compact coordinates and can be
interpreted as solutions of the equations of motion following
from the action of the 4D effective field theory obtained by dimensional
reduction from the 6D action (\ref{F1})\cite{sen}
\begin{equation}
\label{F2}
I_4=\frac{1}{2{\kappa_4}^2}\int d^4 x\sqrt{-G'}e^{-{\p}'}\left [R +
(\partial {\p}' )^2 - \frac{(H')^2}{12}-{1\over4} F(LML)F +
\frac{1}{8}Tr (\partial ML\partial ML)+\ldots \right ].
\end{equation}
Here $G= (G_{mn})$ , $B= (B_{mn}), \quad m,n =1,2$ and  the space-time 
 tensor indices in (\ref{F2}) are raised with respect to the metric
$G'_{\mu\nu}$.
$$   
\begin{array}{ll}
M=\left (
\begin{array}{cc}
\,G^{-1} & G^{-1}B\\
-BG^{-1} & \,G 
\end{array}    
\right ),& \qquad
L=\left (
\begin{array}{ll}
0 & I_2 \\
I_2 & \,0 
\end{array} \right ).
\end{array}
$$

The $4D$ string-frame backgrounds are related to the fields in the Lagrangian 
(\ref{E2}) as \cite{sen}
\begin{eqnarray}
\label{E4}
G'_{\mu\nu} = G_{\mu\nu} - G_{mn} A^{(1)m}_{\mu} A^{(1)n}_{\nu}, \quad
B'_{\mu\nu} = B_{\mu\nu} - B_{mn} A^{(1)m}_{\mu} A^{(1)n}_{\nu}, \nonumber \\
H'_{\m\n\l} = H_{\m\n\l} -(A^{(1)n}H_{n\n\l} -A^{(1)m}A^{(1)n}H_{mn\l} +
cycl. perms.).
\end{eqnarray}
The gauge fields and dilaton are
\begin{equation}
A^{(1)n}_{\mu}  =  G^{nm} G_{m\mu} \ , \ \ \
A^{(2)}_{n\mu } =  B_{n\mu}, \ \ \
{\p}' = {\Phi}'  -{1\over 2}\ln G, \ \ \  G\equiv \det G_{mn},
\label{E5}
\end{equation}
and their explicit form can be read off from (\ref{E2}):
$$
A^{(1)1}_{\mu} = (0,\ a_s), \ \ \ A^{(1)2}_{\mu} = (K^{-1},\ 0),
\ \ \ A^{(2)}_{1 \mu} = (0,\ b_s), \ \ \ A^{(2)}_{2 \mu} = (F,\ 0), \ \ \
B_{mn}= B_{\m\n} =0.
$$
The vectors $a_s$ and $b_s$ are expressed via the functions $f$ and $k^{-1}$
which are solutions of the harmonic equation.

The dilaton   
${\Phi} '$ in (\ref{F1}) can be split in the sum ${\Phi} + {\Phi}_\infty $,
where at spatial infinity ${\Phi}=0$, and ${\Phi}_\infty $ is a free
parameter, and also $ {\p}' ={\p} + {\p}_\infty $. 

 Written in the Einstein frame, where $g_{\m\n}=
e^{-\p}G'_{\m\n}$, the action (\ref{F2}) is \cite{sen}   
\begin{equation}
\label{P1}
I_4 = \int d^4 x \sqrt{-g}\left[ R
-{1\over2}(\partial {\p} )^2 -{e^{-\p}\over4} \F(LML)\F +
\frac{a_1}{4\sqrt{-g}}\F L\z\F +\frac{1}{8}Tr
(\partial ML\partial ML)+\ldots \right ].
\end{equation}
Here we kept only the non-constant part of the dilaton $\p$ and
introduced the axion
$$\pa_\rho a_1 = -{H'}^{\m\n\l }e^{-2\p}\sqrt{-g}e_{\m\n\l\rho }.$$
A class of solutions  of the equations of motion derived from the action
(\ref{P1}) is \cite{bh2,cvyo} (only non-vanishing backgrounds are presented) 
\begin{eqnarray}
\label{E8}
&{}&ds^2=-{\Lambda}(r)dt^2 +{\Lambda}^{-1}(dr^2+r^2d\Omega_2^2),\nonumber \\
&{}&{\Lambda}^2(r)=FK^{-1}kf^{-1},\nonumber \\
&{}&{\p}=\frac{1}{2}\ln FK^{-1}fk^{-1}, \nonumber \\
&{}&G_{11} =FK, \nonumber \\
&{}&G_{22} =fk,
\end{eqnarray}
where
$$  K=A\left(+\frac{Q_1 }{r}\right),
\quad F^{-1}=B \left(1+\frac{Q_2 }{r}\right), \quad
k^{-1}= a\left(1+\frac{P_1 }{r}\right),\quad
f =b\left(1+\frac{P_2 }{r}\right)
$$
are harmonic functions. Demanding that at  spatial infinity the metric
and dilaton are asymptotic to the Lorentzian metric and unity,
 we have $ABab =1$ and $\frac{AB}{ab}
=1$, which is solved by $AB=1$ and $ab =1$. 
$F^{(1)1}$ and $F^{(2)}_1$ are magnetic and  $F^{(1)2}$ and
$F^{(2)}_2$ are electric field strengths.
 $G_{11} $ and $G_{22}$ are the
nonzero components of the metric of the torus $T^2$ on which
are compactified two dimensions of the six-dimensional theory.
From the  $4D$ point of view, the backgrounds are interpreted as charged
black holes.
\section{Loop-corrected 4D effective field theory.}
Compactification of heterotic string theory on the manifold $K3$  or on 
a suitable orbifold yields $6D$ theory with $ N=1$ supersymmetry. 
Further compactification on a two-torus yields a
$N=2$, $4D$ locally supersymmetric theory.

 To investigate solutions of the loop-corrected equations of motion in 
this theory,  we 
make use of the well-developed technique of the $N=2$ supersymmetric
supergravity \cite{wikalu,dewit,cafpr,andr,crap}.
 The form of $N=2$ supergravity depends on the choice of a
holomorphic
vector bundle $(X^A , F_A )$, where $X^A$ denote both the vector superfields
and the complex scalar components of these superfields. 
The moduli and the vector fields are combined in $N=2$ vector
multiplets.  If a holomorphic vector bundle admits a holomorphic prepotential
$F(X)$, the latter completely defines  dynamics of  $N=2$ supersymmetric
theory.
In particular, in this case $F_A = \pa F/{\pa X^A }$. 

 In the heterotic string compactification the prepotential has no perturbative
loop corrections beyond one loop and is of the form
\cite{afgnt,wikalu,defekoz,kou,hm}
\begin{equation}
\label{M6}
F=-\frac{X^1 X^2 X^3}{X^0} -i {X^0}^2 \e h^{(1)}\left(-i\frac{X^2}{X^0},
-i\frac{X^3}{X^0}\right)+\ldots ,
\end{equation}
where dots stand for contribution of other moduli.

 The moduli space is   
parametrized by complex coordinates $z_i =iy_i$. 
Special coordinates are introduced as
$$ \frac{X^i }{X^0 } = iy_i = (Re\,y_i +ia_i ), \quad X^0 =1.
$$
We use the standard  identification of the tree-level special coordinates 
\begin{eqnarray}
&{}&y_1 =S =i(e^{-\p} +i a_1 ), \nonumber \\ 
&{}&y_2 =T =i(\sqrt{G}+iB_{12}),\nonumber \\
&{}&y_3 =U =i\left(\frac{\sqrt{G} +iG_{12}}{G_{22}}\right).
\label{M10}
\end{eqnarray}
In the case of solution (\ref{E8}), $B_{12} = G_{12} =a_1 =0$.

In the perturbative approach, we neglect  non-perturbative corrections
 to the prepotential of the form $f(e^{-2\pi S}, T,U)$. 
The loop corrections are
accompanied by a  factor  $\e $ and are treated perturbatively.
 
 The
bosonic part of the universal sector of the $N=2$ supersymmetric 
theory written in the standard
form of $N=2$ special geometry
\cite{dewit,cafpr,andr,crap,afr}.
in the holomorphic section with the prepotential is
\begin{equation} 
\label{M11}
I_4 = \int d^4 x \sqrt{-g}\left[-\frac{1}{2} R + i\left(\bar{N}_{IJ}
\F^{-I}\F^{-J}-
N_{IJ}\F^{+I}\F^{+J}\right) + k_{i\bar{j}} \pd y_i \pu \bar{y}_{\bar{j}}
 +\ldots\right]
\end{equation}
Here $N_{IJ}$ are the gauge couplings 
\begin{equation}
\label{M9}
N_{IJ}=\bar{F}_{IJ} +2i\frac{(Im F_{IK} X^K )(Im F_{JL} X^L )}
{(X^I Im F_{IJ} X^J )}, 
\end{equation}
where $F_{IJ} ={\pa}^2 F/{\pa X^I \pa X^J} $ and $F$ is the total
loop-corrected prepotential, and (anti)self-dual field strengths are
$$  \F^{\pm }_{\m \n} 
=\frac{1}{2}( \F_{\m\n} \pm \sqrt{-g} \F^{*}_{\m\n}),
$$
where
$ \F^{*}_{\m\n} ={1\over2}e_{\m\n\r\l}\F^{\r\l}$, and $e_{\m\n\r\l}$ is
the flat antisymmetric tensor.

The vector fields are re-labeled in correspondence with the
moduli with which they form the supermultiplets
\begin{equation}
A^1_\m =\A^0_{\m}, \quad B_{1\m} =\A^1_{\m} ,\quad A^2_{\m} =\A^2_{\m} , 
\quad B_{2\m} =\A^3_{\m}.
\label{M12}
\end{equation}

The Kaehler metric $k_{i\bar{j}}$ is expressed through the Kaehler potential
$$K=-\log i(\bar{X}^A F_A -{X}^A \bar{F}_A).$$ 

The dilaton kinetic term and last term in the tree-level action 
(\ref{P1}) $\frac{1}{8}Tr(\partial ML\partial ML)$ can be written in terms
of the tree-level Kahler potential
$$ K^{(0)} = -\log[(T+\bar{T})(U+\bar{U})(S+\bar{S})
$$
as
$$  k^{(0)}_{i\bar{j}} \pd y_i \pu \bar{y}_{\bar{j}}=
K^{(0)}_{S\bar{S}}\pa S \pa\bar{S}+ K^{(0)}_{T\bar{T}}\pa T \pa \bar{T}+
K^{(0)}_{U\bar{U}}\pa U \pa \bar{U}.
$$
The loop-corrected Kaehler potential  calculated by using the
loop-corrected prepotential (\ref{M6})  
 is of the form \cite{afgnt,wikalu,kou}
\begin{equation}
\label{M7}
K = -\log[(T+\bar{T})(U+\bar{U})(S+\bar{S}
+\e V(T,\bar{T},U,\bar{U})],
\end{equation}
where the Green-Schwarz function $V$ is 
\begin{equation}
\label{M8}
V= \frac{Re\,h^{(1)} -Re\,T Re\,\pa_T h^{(1)}-Re\,U Re\,\pa_U h^{(1)}}{Re\,T
Re\,U}.
\end{equation}
Beyond the tree level the dilaton mixes with other moduli. At the one-loop
level
\begin{equation}
\label{c1} 
S= e^{-\phi} -\frac{V}{2} + ia_1 .
\end{equation}

The kinetic terms for the moduli calculated with the prepotential (\ref{M6})
are
\be
\label{L12}
4\frac{\pa^2 K}{\pa y_i \pa \bar{y}_{\bar{j}}}
\pd z^i \pu \bar{z}^{\bar{j}}
=(\pa\p )^2 + \frac{|\pa y_a |^2}{(Re y_a )^2}  +
e^{2\p}(\pa_\m a_1 + X_\m )^2 
+\frac{e^\p}{2}V_{y_a\bar{y}_b}\pa y_a \pa\bar{y}_b . 
\ee

Here $$ iX_\m=\frac{1}{2}(V_{\bar{y}_a} \pa_\m\bar{y}_a 
 - V_{y_a} \pa_m y_a ).$$

The loop-corrected universal part of the heterotic string effective action
can be written as
\ba
\label{B3}
I_4 = \int d^4 x \sqrt{-g}\left[-\frac{1}{2} R + i\left(\bar{N}_{IJ}
\F^{-I}\F^{-J}-
N_{IJ}\F^{+I}\F^{+J}\right) + \frac{1}{4}\left((\pa\p )^2 
+e^{2\p}(\pa a_1 +X)^2 \right)\right.
\\\nonumber\left.
+\frac{|\pa T|^2}{(Re T)^2} + \frac{|\pa U|^2}{(Re U)^2}
+\frac{e^\p}{2}\left(V_{T\bar{T}}|\pa T|^2 +V_{U\bar{U}}|\pa U|^2 +
V_{U\bar{T}}\pa U\pa\bar{U} +V_{T\bar{U}}\pa T\pa\bar{U}\right)
 +\ldots\right]
\ea

Starting with the compactified heterotic string theory, one naturally 
arrives at the effective action written in terms of the "heterotic" 
 vector bundle. 
In this holomorphic section the moduli are treated non-symmetrically, 
the modulus
$S$ playing the role of the coupling constant \cite{wikalu,duff,bkrsw}, 
and the vector
couplings become weak in the large-dilaton limit.
It is known that in terms of the heterotic vector bundle it is
impossible to introduce a prepotential, however, 
one can calculate the gauge couplings in this 
holomorphic section by performing a symplectic transformation from the section
with the prepotential \cite{wikalu,cafpr}. 
Following \cite{wikalu}, the heterotic holomorphic section
  $(\hat{X}, \hat{F})$ is obtained from that admitting the 
prepotential by performing  symplectic transformation
$$
\left(\begin{array}{c}
\hat{X}\\ \hat{F}\end{array}\right) =O\left(\begin{array}{c}
X\\ F\end{array}\right)\qquad O=
\left(\begin{array}{cc}
U &Z\\W &V \end{array}\right),  
$$
where the nonzero elements of the matrix $O$ are $U^I_{\,\,J}
=V_{J}^{\,\,I}=\delta^I_J$ for $I,J \neq 1$, $W_{11} =-Z^{11}=\pm 1$.
The gauge coupling constants transform as
\begin{equation}
\label{B7}
 \hat{N} = (VN+W)(U+ZN)^{-1}.
\end {equation}
\section{Equations of motion}
We look for a static spherically-symmetric solution to the equations of
motion derived from the loop-corrected effective action with the final goal
to find the loop-corrected dilaton and metric. In magnetic case, 
the metric and dilaton have singular
behavior at the origin, while other background fields stay constant, and
only for the former we can expect  qualitative effects of loop
corrections.  

We shall concentrate on the case of tree-level solutions which are 
 purely magnetic 
$( Q_1 =Q_2 =0)$ extremal black holes with equal magnetic charges 
$P_1 =P_2 =P$. These solutions can be embedded in $4D\,\, N=2$ dilatonic 
supergravity and leave $1/2$ of the supersymmetry unbroken.
In this case, the non-vanishing backgrounds of the
chiral null model are expressed via a single function (we consider
one-center solution) $ f_{0}$:
\begin{eqnarray}
\label{E9}
ds^2 = -{\f0}^{-1}dt^2 + \f0 {dx^i}^2,\quad f_{0}(r)=1+\frac{P}{r}, \nonumber \\
\p =\ln \f0, \quad a_\varphi  =a^{-1}P(1-\cos\vartheta),\quad 
b_\varphi =aP(1-\cos\vartheta),
\end{eqnarray}
where $a_\varphi$ and $b_\varphi$ are nonzero components of potentials in
spherical coordinates.
The tree-level components of the internal metric $G_{mn}$ are
\begin{equation}
\label{B1}
G_{11} =a^2 , \quad G_{22} =A^2 .
\end {equation}
 The magnetic field strengths are
\begin{equation}
\label{B2}
F^{(1)1}_{ij} =a^{-1}F_{ij}, \quad F^{(2)}_{1ij} =aF_{ij}, \quad F_{ij} =
-\varepsilon_{ijk}\pa^k \f0, 
\end {equation}
 and in spherical coordinates have a single
 nonzero component $F_{\vartheta\varphi}= P\sin\vartheta $.

At the tree level, in the case of magnetic black
hole solution, the imaginary parts of the  moduli, $a_i$,
vanish, and, if appear at the one-loop level, are of order $O(\e )$. 
With the accuracy of the terms of higher                
order in $O({\e})$, the imaginary parts of the moduli $S,\, T$ and $U$
do not enter the equations for the real parts.

It is convenient to introduce the functions $\g$ and $\s$ as
$$ Re T=e^{\g + \s},\quad Re U=e^{\g - \s}.$$         
We have $a =e^{\g_{(0)}},\,A=e^{\s_{(0)}}$.
At the tree level, the functions $\g_{(0)}$ and $\s_{(0)}$ are constants,
and $\pa\g  ,\,\, \pa\s $ are $ O(\e)$.                                       
                                                      
Let us consider the gauge part of the action in the holomorphic section
associated with the heterotic string compactification. The gauge terms can
be also written as
\begin{equation}
\label{L1}
 Im\, \h{N}_{IJ}\, \h{\F}^{I}_{\mu \nu }\h{\F}^{\mu \nu J}
+\frac{1}{\sqrt{-g}}Re \,\h{N}_{IJ}\,
\h{\F}^{* I}_{\mu\nu }\h{\F}^{ \mu\nu J}.
\end{equation}
 The gauge
couplings we need are calculated using (\ref{M9}) and the full prepotential
(\ref{M6}) and  are presented in Appendix B.

Since at the one-loop level only magnetic fields are
present, the second term in (\ref{L1}) is zero. At the one-loop level, this
term can appear only if the  electric fields with the charges of
order $O(\e )$ are generated. Since only the couplings $ N_{0i} $
contain real parts and these are of order $O(\e )$,  the
(topological) terms proportional to $Re N_{0i}$ are of order $O(\e^2 )$.
To conclude, all the terms in the gauge part of the action,
 which are absent at the tree level, at the one-loop level are of order
$O(\e^2 )$
\footnote{This refers also to  electric fields
 which can appear with the charges of order $O(\e )$, see below.}.

At the tree level, only the couplings $\h{N}_{II}$  are non-vanishing. 
In the case of magnetic black hole, the relevant tree-level couplings in
the "heterotic"  basis associated with the heterotic
string compactification are
$$\hat{N}^{(0)}_{00} = STU =e^{-\p}G_{11}, \qquad 
\hat{N}^{(0)}_{11} = S(TU)^{-1} = e^{-\p}G^{11},$$
and are equal to those in the action (\ref{P1}).
At the one-loop level, there appear non-diagonal couplings and
 $O(\e )$ corrections to the diagonal 
couplings (see Appendix B).

The loop-corrected gauge term $ Im\,
\h{N}_{IJ}\, \h{\F}^{I}_{\mu \nu }\h{\F}^{\mu \nu J}$ 
 written explicitly  with the gauge couplings $\h{N}_{IJ}$  
presented in Appendix B, is  
\begin{eqnarray}
\label{B4}
L_g =
-2\left[\left(e^{-\p +2\g } -\e\frac{n+2v}{4}\right)
({\h{\F}}^{0})^2 +
\left(e^{-\p-2\g } -\e\frac{n+2v}{4}e^{-4\g}\right) (\h{\F}^{1})^2 
+\e\frac{n+2v}{2}e^{-2\g}
 (\h{\F}^{0}\h{\F}^{1})\right]
\end {eqnarray}
The functions $v =TU\,V(T, U)$ and  $n$ are  defined in Appendix B.
Identification of the fields in (\ref{B3}) and (\ref{B4}) is as in
(\ref{M12}): 
\begin{equation}
\label{M14}
F^{(1)1} =\sqrt{8}\h{\F}^0 ,\qquad F_1^{(2)}=\sqrt{8}\h{\F}^1 ,
\end{equation}
the factor $\sqrt{8}$ is due to different normalizations of the actions.
  
General ansatz for the spherically-symmetric metric is 
\begin{equation}
\label{E12}
ds_4^2 =- e^{\nu} dt^2 + e^{\l}dr^2 + e^{\mu}d\Omega_2^2.
\end{equation}

In the leading order, from (\ref{E9}) the metric components, dilaton and
moduli are
\begin{eqnarray}
\label{E24}
{\nu}_{(0)} &=&-\ln\f0,\quad \l_{(0)} =\ln\f0,\quad \mu_{(0)} =\ln\f0 +2\ln r,
\quad \p_{(0)} = \ln\f0, \nonumber \\
e^{\g_{(0)}} &\equiv& e^{\g_0} =aA,\quad e^{\s_{(0)}} \equiv e^{\s_0}=\frac{a}{A}.
\end{eqnarray}
The tree-level field strengths are those in (\ref{B2}), and
$$ {F^2}_{(0)} =2{q'}^2, \quad q' \equiv\frac{\f0'}{\f0}.$$ 
In the first order in  $\e$, we look for a solution in the form
\begin{eqnarray}
\nu &=&-\ln\f0 +\e n ,\quad \l =\ln\f0 +\e l ,\quad \mu =\ln\f0 +2\ln r +\e m,
\quad \p = \ln\f0 + \e\varphi, \nonumber \\
\g &=&\g_0 +\e\g_1 , \quad \s =\s_0 +\e\s_1 .
\label{E25}
\end{eqnarray}
Here $n,m,l,\p , \g_1 $ and $\s_1 $ are unknown functions which are 
determined from the field equations.

The system of Maxwell equations and  Bianchi identities for the 
gauge field strengths obtained from the action (\ref{L1}) is
\begin{equation}
\label{L3}
\pd \left(\sqrt{-g} Im \hat{N}_{IJ} \hat{\F}^J + Re \hat{N}_{IJ} 
\hat{\F}^{*J}\right)^{\m\n} =0
\end{equation}
and
\begin{equation} 
\label{L4}
\pd \h{\F}^{*J\m\n} =0
\end{equation}
Let us consider Eq.(\ref{L3}) with $I=0$. The $\n =0$ component of this
equation is
\begin{equation}
\label{L5}
\partial_r  \left(\sqrt{-g} Im\h{N}_{0I} \F^I + Re \h{N}_{0I} \h{\F}^{I}\right)^{0r} 
=0 .
\end{equation}
At the tree level, $\h{\F}^{0,1\, 0r} =0$, and (\ref{L5}) is satisfied
identically. At the one-loop level, noting that  $Im \h{N}_{00} = 
O(1)$ and $Im \h{N}_{0i} =O(\e )$, keeping only the terms of the highest
order in $\e$, we
obtain solution of (\ref{L5}) in the form
$$ \h{\F}^{0\, 0r} =\frac{\e c_0 (\vartheta,\varphi)}{\sqrt{-g'} 
Im \h{N}_{00}}, $$
where $-g'(r) =e^{\n +\l +2\m}$ and $c_0 (\vartheta,\varphi)$ is an arbitrary
function.  Bianchi identity (\ref{L4}) shows that $c_0 =Const$. In the
same way, the equation with $I=1$ yields
$$\h{\F}^{0\, 1r} =\frac{\e c_1}{\sqrt{-g'} Im \h{N}_{11}}. $$

The $\n =\vartheta $ component of Eq.(\ref{L3}) with $I=0$ (the 
$\varphi$
component yields the same result) with the accuracy of the terms of order
$O(\e)$ is
$$\pa_{\vartheta}  \left(\sqrt{-g} Im \h{N}_{00} \h{\F}^0 + Im \h{N}_{01}
\h{\F}^{1}\right)^{\vartheta\varphi} =0, $$
and for spherically-symmetric fields is satisfied identically.  
Here we kept only the 
terms of the leading and the first orders in $\e$. Similar equation holds 
for $I=1$. Bianchi identities are 
$$ \pa_r \h{\F}^{0,1}_{\vartheta\varphi}=0,  $$ 
implying that the field strengths have the form
$$\h{\F}_{\varphi\vartheta}^{0,1} =P^{0,1}\sin \vartheta.$$ 
Comparing with the field strengths (\ref{B2}) and (\ref{M14}), we find that
\begin{equation}
\label{M13}
P^0 =\frac{e^{-\g_0 }P}{\sqrt{8}}, \quad P^1 =\frac{e^{\g_0 }P}{\sqrt{8}}.  
\end{equation}
The fields $\h{\F}^{0\, 0r}$ and $\h{\F}^{1\, 0r}$,
being of order $O(\e )$, contribute to the
Einstein and dilaton equations  the terms of order $O(\e^2 )$.

Keeping in the dilaton  equation of motion the terms up to the order
$O(\e)$, we obtain  
\begin{eqnarray}
\label{B5}
\frac{1}{\sqrt{-g}}\pa_{\mu }\left(g^{\mu\nu }\sqrt{-g}\pa_{\nu }
\p \right)
 +  \frac{1}{4} \left(e^{-\p+2\g} ( F^{(1)1})^2 
+e^{-\p-2\g}(F_1^{(2)})^2\right)=0.
\end{eqnarray}
The terms with $\pa\s$ and $\pa\g$ are of the next order on string coupling
and are omitted.
 With the required accuracy, the terms with the gauge fields can be written as
$${1\over 4}e^{-\p}\left[e^{2\g_0 }(1+2\e\g_1)( F^{(1)1})^2 +e^{2\g_0
}(1-2\e\g_1)(F_1^{(2)})^2 \right].$$
In view of (\ref{M13}), the terms with $\g_1$ cancel, and we are left with
\begin{eqnarray}
\label{E14}
\frac{1}{\sqrt{-g}}\pa_{\mu }\left(g^{\mu\nu }\sqrt{-g}\,\pa_{\nu }
\p \right)
 + \frac{1}{2} e^{-\p }F^2  =0.
\end{eqnarray}

With the required accuracy the Einstein equations can be written as
\begin{equation}
\label{E16}
R_{\mu\nu}-\frac{1}{2}g_{\mu\nu}R -\frac{1}{2} 
\left(\pd\p \pa_{\nu}\p -\frac{1}{2}g_{\mu\nu}(\pa\p)^2 \right)
+ {(L_g)}_{\m\n} -{1\over 2} g_{\m\n}L_g =0.
\end{equation}
Here
$$ L_g = {1\over 2}e^{-\p}  F^2, \qquad
{(L_g)}_{\m\n}={1\over 2}e^{-\p}(F^2)_{\m\n}. 
$$
The field strengths squared have the following nonzero components
$$
(F^2)_{\varphi}^{\varphi} =
(F^2)_{\vartheta}^{\vartheta} = \frac{F^2 }{2},$$
where
\begin{equation}
\label{E18}
F^2 =2{q'}^2 (1-2\e m)).
\end{equation}

The functions $\s_1$ and $\g_1$ decouple from the above equations.
Since in this paper we are interested in the form of loop-corrected 4D metric
and dilaton, 
we shall not determine these functions explicitly ( the corresponding
equations of motion are solved elsewhere \cite{mi} ).

The Einstein equations (\ref{E16}) (with one index lifted) take the form
\cite{ll}:
\begin{equation}
\label{E21}
e^{-\l}\left(\mu'' + \frac{3}{4}{\mu'}^2 -\frac{\mu'\l'}{2}\right)-e^{-\mu}+
\frac{1}{4}e^{-\l}{\p'}^2 +
\frac{1}{4}e^{-\phi}F^2 =0,
\end{equation}
\begin{equation}
\label{E22}
e^{-\l}\left(\frac{{\mu'}^2}{2}
+\mu'\nu'\right)-2e^{-\mu}-\frac{1}{2}e^{-\l}{\p'}^2 +
\frac{1}{2}e^{-\p}F^2=0,
\end{equation}                                                   
\begin{equation}                                                 
\label{E23}          
e^{-\l}(2\mu''+2\nu''+{\mu'}^2+{\nu'}^2-\mu'\l'-\nu'\l'+\mu'\nu')
+e^{-\l}{\p'}^2 -e^{-\phi}F^2=0.
\end{equation}             
With the required accuracy, in the equations of motion, all the expressions
of the first order in string coupling $\e$ are 
calculated with the tree-level moduli.
\section{Loop-corrected metric and dilaton}
To solve the loop-corrected equations of motion with the required accuracy,
i.e. in the first order in the loop-counting parameter $\e$, we substitute
in the above equations the functions $\m, \n, \l$ and $\p$ in the form
(\ref{E25}) 
In Appendix A we present the detailed solution of the equations of motion 
in the ansatz $ l=-n$ 
\footnote{There are four
equations (\ref{E14}) and (\ref{E21})-(\ref{E23}) for three unknown functions
$m,\, n $ and $\varphi$.  Our choice corresponds to the
requirement that in the first order in $\e$, as in the leading order, $\nu =
-\l$.}. 

\begin{eqnarray}
n=\frac{P}{r}(A_1 -\frac{C_2}{2}) - A_2\frac{P}{r+P}\qquad
m=-\frac{P}{r}A_1  + A_2\frac{P}{r+P}\qquad
\label{E26}
\end{eqnarray}
and
\begin{equation} 
\label{E27}
\varphi=\frac{P}{r}A_1 +A_2\frac{P}{r+P}.
\end{equation}  
Here $A_1$, $A_2$ and $C_2$ are arbitrary constants.

By coordinate transformation the metric (\ref{E12}) with $\n =-\l$ can be
reduced to the form
\begin{equation}
\label{E28}
 ds^2 =-e^{U(R)} dt^2 + e^{-U(R)} (dR^2 + R^2 d{\Omega}^2),	  
\end{equation}
where the new variable $R$ is determined from the relation 
$$\ln\frac{R}{R_1} =\int^r_{r_1}\, dr' e^{\frac{1}{2}(\l -\mu )(r')}. $$
For the  solution (\ref{E26}), we have $R=C e^{\frac{\e C_2}{2 r}}$,
where $C$ is an arbitrary constant. In new coordinates, the asymptotics of
the metric and dilaton are obtained by setting in (\ref{E26}) $C_2 =0$.   

The ADM mass calculated with the metric (\ref{E28}) is
$$ M=2P(1+\e A_1 ).
$$

The expressions (\ref{E26})-(\ref{E27}) were
obtained by making the expansions of
$e^{\e n},\,e^{\e l}$, etc. to the first order in $\e$ assuming that
$1>|\e n |,\,|\e l |, \ldots$. This yields
$$r>\e P(|A_1 |+ |A_2 |).
$$

Solving the system of the Einstein-Maxwell equations and the
equations of motion for the moduli, we cannot decide
which solution is supersymmetric. This issue can be solved by studying the
$N=2$ supersymmetry transformations and solving Killing spinor equations
 \cite{mi}.  
We found that the supersymmetric solution with the metric of the
form (\ref{E28})  corresponds to
\begin{equation}
\label{k5}
A_1 = P|h|TU/2
\end{equation}
(in notations of Sect.4 $TU =G_{11}= e^{2\g_{0}}=a^2$).   
From the explicit form of the prepotential which was calculated
in \cite{hm} for the
case of unbroken gauge group is $[E_8\times E_7\times U(1)^2 ]_L
\times U(1)^2_R $ follows that $h$ is negative 
\footnote{Strictly speaking, in the above expressions for the gauge 
couplings above stands $Re\,h$. Prepotential
$h$ is defined up to a quadratic polynomial in $T,\,U$ and $TU$ with
imaginary coefficients. Although the gauge couplings contain the
ambiguities, these cancel in the expressions for the field strengths, and
the final results are unambiguous \cite{mi}}.
 and the loop-corrected metric is
 \label{D2}
\begin{equation}
g^{ii}= -g_{00} = \frac{r}{r+P}\left(1+\e\frac{P|h|/2TU}{r}\right).
\end{equation}
Extrapolating this expression to the region of of small $r$, 
 we note that
 the singularity at $r=0$ is smeared by quantum corrections. 
\section{Discussion}

In this paper we discussed solutions to the equations of motion of the
string-loop-corrected $4D, \quad N=2$ effective action obtained by 
dimensional reduction on a two-torus from the $6D,  \quad N=1$ effective action 
of heterotic string theory. The effective action contains both the dilaton and
 the metric of compact dimensions. As a tree-level solution, we
considered the magnetic black hole solution.  

Different embeddings of $N=2\, \,STU$ model in type IIA and IIB theories 
and construction of the generating solution of regular BPS tree-level 
black hole solutions were studied in \cite{bt} and refs. therein. 
Embeddings with charges stemming from R-R and NS-NS sectors are possible.
In the case of heterotic embedding considered in the present paper the 
generating solution is the same as in the case of non-symmetric NS-NS
embedding of the $STU$ model in toroidally compactified type II theory.
From the point of view of the perturbation theory, the above embedding is 
singled out, because in this case dilaton is one of the moduli, and the 
dilaton-axion parametrize a separate factor of the moduli space.
   
Our discussion was  simplified by considering the tree-level solution
with  equal magnetic charges,
in which case the string-loop corrections are independent of coordinates, 
although our treatment can be extended to the case of unequal charges.

Let us discuss modification of the above results in the case of unequal
charges $P_1 \neq P_2$
(for simplicity we set in (\ref{B1}) $a=A=1$). We are mainly interested in
the small-$r$ region. The tree-level expressions are modified in the following
way: the functions $\m_{(0)} , \n_{(0)}, \l_{(0)}$ and $\p_{(0)}$ 
retain the same functional form
(\ref{E24}) as at the tree level, except for the function $\f0$ which is
changed to
\begin{equation}
\label{AD1}
 \tilde{f_0 } =\left[\left(1+{P_1\over r}\right)\left(1+{P_2\over r}
\right)\right]^{1/2}. 
\end{equation}
The squares of field strengths are 
$$ {F^{(i)}}^2 =2P_i^2\frac{1}{{\tilde{f_0 }}^2 r^4}.$$
For the metric components of the internal two-torus we obtain 
$$G_{11} = \frac{P_1 +r}{P_2 +r} = \frac {P_1}{P_2} +r\left({1\over P_2}
-{1\over P_1}\right) +O(r^2), \quad G_{22}=1 .
$$
The modulus $\g$ is now $r$-dependent, and at small $r$ we have 
\begin{equation}
\label{AD2}
\pa_r\g = a+\e \pa_r\g_{1}, \quad a=\frac{1}{2}\left({1\over P_2}
-{1\over P_1}\right).
\end{equation}
Except for a substitution of $\f0$ by $\tilde{f_0 }$, in the gauge action
 there appear new terms of order $O(\e )$ 
$$(\pa\g)^2 \e e^\p F_{\g\g}+2\pa\p \pa\g \e e^\p F_{\p\g},$$
where the functions $F_{\g\g}$ and $F_{\p\g}$ are expressed via the
derivatives of the prepotential.   

In the limit $r\rightarrow 0$, the $O(\e)$ terms with the most singular
 behavior stemming from the gauge and dilaton parts of the action are 
$$ \e{v\over 2}e^{-2\g} P_1 P_2 e^{-\p} \frac{1}{{\tilde{f_0}}^2 r^4}$$
and 
$$ \e (\pa\p)^2 v e^{-2\g}e^\p = -\e {\tilde {q'}}^2 v e^{-2\g}e^\p,$$
where $\tilde
{q'}=\frac{\tilde{f'_0}}{\tilde{f_0}}$.
At small $r$, these terms retain the same functional form as in the case $P_1 =P_2$,
 because ${\tilde {q'}}^2 =\frac{P_1
P_2}{{\tilde{f_0}}^2 r^4} (1+O(r^2))$.

As a result, in the case of unequal charges the  equations of motion retain
 the same functional form as for equal charges,
except for the extra terms  of order
$\e a\tilde{q'}e^\p$, where $a$ is defined in (\ref{AD2}).
 Solving the equations of motion along the lines
of Appendix 1, we find that solutions obtained in the case of equal charges
 are modified by
terms of order $O(a\ln r)$ which at small $r$ are much smaller than the
leading terms of order $\frac{1}{r}$. 

Another way to obtain the loop-corrected expressions for the metric and
moduli is to solve the spinor Killing equations for the $N=2$ supersymmetric 
loop-corrected effective action. Details of this work are reported
elsewhere \cite{mi}.

Perturbative expansion in string coupling implies that the non-perturbative
corrections of the form $O(e^{-2\pi S})$ are absent. This means that in the
weak coupling limit we consider, we cannot check duality properties of the 
full theory which are restored only in the full non-perturbative setting
\cite{nper}.
 
The problem of string-loop corrections to the classical charged black holes
in the effective $N=2$ supergravity was studied also in papers
\cite{bls,bgl}.
However, the two approaches are rather different. In the these papers, the loop
corrections were calculated under an assumption that there exists a "small"
modulus and it is possible to expand the loop correction to the 
prepotential with respect to the ratios
of small to large moduli. Within the framework of string perturbation
theory,
 the natural  expansion  parameter is connected with the  dilaton, 
but this modulus
does not enter the loop correction to the prepotential. The remaining two
moduli connected with the metric of the compact two-torus for special
configurations may have
parametric smallness, but not the functional one connected with dependence
on $r$. Moreover, as we have
argued above, to study the loop corrected solution it is important to take into
account  corrections to the gauge couplings.

 The string-tree-level chiral null model provides a  solution to
 the low energy effective action  which, in a special
renormalization  scheme,  receives no
$\a'$ corrections \cite{bh2,howe}. The loop-corrected solution  
 is no longer expressed in terms of  harmonic functions, and the
$\a'$-corrections are present.
However, still it is possible that $\a'$ corrections are small and can 
be treated perturbatively.
It may be noted here that smearing of the singularity of the point-like
source by $\alpha'$-corrections was discussed previously in \cite{tssm}.

{\large\bf Acknowledgments}

I would like to thank R. Kallosh for helpful
correspondence, A. Marshakov, O. Kechkin, A. Sagnotti and I.Tyutin for remarks and
discussion.

The work of M.I. was  partially supported by the RFFR  grant No 00-02-17679.
\appendix
\section{Solution of field equations}
In the first order in parameter $\e$ the Einstein equations
(\ref{E21})-(\ref{E23}) are \footnote{Below we work with dimensionless
variable $r$. In the final expressions $P$ is reinstated by substitution
$r\rightarrow {r\over P}$.}
\begin{equation}
\label{A1}
m''+m'(q'+\frac{3}{r})-l'(\frac{1}{2}q'+\frac{1}{r})
+{\vp}'\frac{q'}{2} -\frac{l-m}{r^2}+\frac{1}{2}{q'}^2 s  =0,
\end{equation}
\begin{equation}
\label{A2}
m'\frac{2}{r}+n'(q'+\frac{2}{r})-{\vp}'q'-2\frac{l-m}{r^2}+{q'}^2 s  =0,
\end{equation}
\begin{equation}
\label{A3}
m''+n'' +m'\frac{2}{r} -l'\frac{1}{r} + n'(-q' +\frac{1}{r}) +
{\vp}'q' -{q'}^2 s  =0.
\end{equation}
Here 
\begin{equation}
\label{A4}
s =l-\vp -2m 
\end{equation}
We also need the equation for the dilaton (\ref{E14}) in
the  $O(\e)$ order:
\begin{equation}
\label{A5}
{\vp}''+\frac{2}{r}{\vp}'+\frac{1}{2}(2m'+n'-l')q' +{q'}^2 s +a\f0 {q'}^2=0.
\end{equation}
We look for a solution such that $l=-n$, because in this case, as at the tree
level, the components of the metric satisfy the relation $g_{tt} =
 g^{-1}_{rr}$. Substituting this ansatz in the above equations
and forming the combination of equations $(\ref{A1})-\frac{1}{2}(\ref{A2})$,  
we have
\begin{equation}
\label{A6}
m''+m'(q'+\frac{2}{r}) +{\vp}'q' =0.
\end{equation}
Taking into account the explicit form of the functions $f_0$ and
$q'=\frac{{f_0}'}{f_0}$, we integrate this equation and obtain the first
integral as
\begin{equation}
\label{A7}
m'+q'(C_1 +{\vp}) =0,
\end{equation}
where $C_1$ is an integration constant.
To solve Eqs.(\ref{A2}) and (\ref{A3}) we introduce the combinations $u=m+n$
and $t=n-\vp$. The function $s$ in (\ref{A4}) is
$$
s=-2u+t.
$$
Esq.(\ref{A2}) and (\ref{A3}) take the form
\begin{eqnarray}
\label{A8}
u'\frac{2}{r}+ t'q'+2\frac{u}{r^2} + {q'}^2 s  =0,
\nonumber \\
u'' +u'\frac{2}{r}- t'q' - {q'}^2 s  =0.
\end{eqnarray}
The sum of these equations is
\begin{equation}
\label{A9}
u'' +u'\frac{4}{r} +u\frac{2}{r^2} =0
\end{equation}
yielding  solution $u=\frac{C_2}{r} +\frac{{C_2}'}{r^2}$. We set
${C_2}'=0$ to avoid a solution with a too rapid growth of the 
function at small $r$.
Substituting $u=\frac{C_2}{r}$ in Eq. (\ref{A8}) (either equation yields the
same result), we arrive at the equation
\begin{equation}
\label{A10}
t' +q't = q'\frac{2C_2}{r} .
\end{equation}
with  the solution
\begin{equation}
\label{A11}
t=C_3 +\frac{C_2 }{r} -\frac{C_2+ C_3 }{r+1}.
\end{equation}
Substituting $\vp = -m+u-t$ in  (\ref{A7}), we solve this equation and
obtain
\begin{eqnarray}
\label{A12}
m= - \frac{1}{r}(C_1
+\frac{C_2}{2}-\frac{C_3}{2}-C_4 )  
+\frac{1}{r+1}({C_2\over 2}+{C_3\over 2}) +C_4.
\end{eqnarray}
Using (\ref{A11}) for $t$,  (\ref{A12}) for $m$, and $u=\frac{C_2}{r}$,
 we have
\begin{eqnarray}
\label{A13}
\vp = \frac{1}{r}(C_1
+\frac{C_2}{2}-\frac{C_3}{2}-C_4 ) 
+\frac{1}{(r+1)}({C_2\over 2}  +{C_3\over 2} )-C_3 -C_4.
\end{eqnarray}            
Eq. (A5) with $m+n = u = \frac{C_2}{r}$ and $s=-2u+t$ with $t$ from
(A11) yields
\begin{eqnarray}
\label{A14}
{\vp}' =  
({C_2\over 2}  +{C_3\over 2})\left(\frac{1}{r^2 } -\frac{1}{(r+1)^2}
\right) +\frac{C_5}{r^2}.
\end{eqnarray}
Comparing (\ref{A14}) with the derivative of the solution (\ref{A13}), we
find that the coefficients at the terms $(r+1)^{-2}$ coincide, and
coefficients at the terms $r^{-2}$ yield the relation expressing the
constant $C_5$ in terms of other constants.

Next we require that  in the limit
$r\rightarrow\infty$ our solutions should be asymptotic to the Lorentzian
metric. This requirement sets $C_3 =C_4 =0$ and we are left with
\begin{eqnarray}
\label{A15}
\vp &=& \frac{1}{r}(C_1 +\frac{C_2}{2})+
\frac{1}{r+1}{C_2 \over 2}, 
\nonumber \\
m &=&  - \frac{1}{r}(C_1 +\frac{C_2}{2})
 +\frac{1}{r+1}{C_2 \over 2} , \nonumber \\
n &=& \frac{1}{r}(C_1 -\frac{C_2}{2})-\frac{1}{r+1}{C_2 \over
2}.
\end{eqnarray}
Introducing new constants 
$$ A_1 =C_1-\frac{C_2}{2}  $$
and
$$A_2 = {C_2 \over 2} ,$$
we obtain the expressions discussed in Sect.5.

\section{Loop-corrected gauge couplings}

We remind the notations: $\frac{X^i}{X^0} =iy_i$, where 
$$ (y_1 , y_2 , y_3)=(S,\,T,\,U)$$
and consider real $S, T, U$ relevant to this paper.
The prepotential $ {\cal F} =-i{X^0}^2 F(X)$ is 
$$ {\cal F}(z) = y_1 y_2 y_3 +\e h(y_2 , y_3 ), $$
where $h(y_1 , y_2 )$ is a real function. Define $h_i
\partial y_j$ and
$$v= h-y_i h_i,$$
$$n= h- y_i h_i +y_i h_{ij} y_j.
$$
The function $v$ is connected with the Green-Schwarz function $V$ introduced
in Sect.3 as
$$ v(T, U) =V(T, U)TU = h^{(1)}-T\pa_T h^{(1)} -U\pa_U h^{(1)}.$$ 
Because the functions $V$ and $v$ appear in the first order in the string
coupling  $\e$, they are calculated with the tree-level moduli. In particular, 
we have  $V= ve^{-2\g_0}$. 

In the basis admitting a prepotential, the gauge couplings are calculated using the
formula (\ref{M9}), and we have (we list only those used in our calculations) 
\begin{eqnarray}
\label{C1}
N_{00}& =&-i\left(y_1 y_2 y_3 -\e \frac{n}{4}\right), \nonumber \\
N_{01}& =&-\e\frac{n +2v}{4y_1 } +i\e a_1\frac{y_2 y_3}{y_1 }, \nonumber \\
N_{11}& =&-i\frac{y_2  y_3 }{y_1 }\left(1 +\e\frac{n}{4y_1 y_2
y_3}\right).
\end{eqnarray}
The couplings in the "heterotic" basis are obtained via the transformation
(\ref{B7}). In the first order in $\e$ we obtain
\begin{eqnarray}
\label{C2}
\hat{N}_{00}& = &N_{00} =-iy_1 y_2 y_3 \left(1 -\e \frac{n}{4y_1 y_2
y_3}\right) =-i\left(e^{-\p +2\g} -\e \frac{n}{4} \right), \nonumber \\
\hat{N}_{01}& =&-\frac{N_{01}}{N_{11}}= -i\e\frac{n+2v}{y_2 y_3 } +\e a_1=
i\e e^{-2\g}(n+2v) +\e a_1 , \nonumber \\
\hat{N}_{11}& =&-\frac{1}{N_{11}} =-i\frac{y_1 }{y_2  y_3 }
\left(1 -\e\frac{n}{4y_1 y_2 y_3}\right)=-i\left(e^{-\p -2\g}
-\e\frac{n}{4}e^{-4\g} \right).
\end{eqnarray}

The Green-Schwarz function $V(T,\bar{T},U,\bar{U})$ was
estimated numerically  in \cite{kou} and was shown to be positive. 
Note also that for small $T-U$, i.e. near the points
of enhanced symmetry $T=U$, $ h^{(1)}(T,U) = -\frac{1}{16\pi^2
}(T-U)^2\log(T-U)^2$ + regular, where the regular term is finite for
$T\approx U\neq 1, e^{\pi i/6}$. The second derivatives of the function
$h^{(1)}(T,U)$ at the points of the enhanced symmetry have logarithmic
singularities. To avoid these points, we introduced in (\ref{B1}) the
constants $a$ and $A$ choosing them so that they are not connected by a
rational function thus  ensuring  that the moduli cannot be transformed to
 each other by a modular transformation. However, all final expressions
depend only on the function $V$, which contains only the first derivatives
of the prepotential and is non-singular for all values of $T$ and $U$.
Thus, our expressions are valid for any $a$ and $A$.

\vskip2.mm

\end{document}